\documentclass[manuscript,showkeys,amsmath,amssymb,12pt]{revtex4-1}
\usepackage[T1]{fontenc}
\usepackage{color}
\usepackage[latin1]{inputenc}
\usepackage{graphicx}
\usepackage{epsfig}
\usepackage{rotating}
\usepackage{dcolumn}
\usepackage{bm}

\begin{document}

\title{Anisotropic line tension of lipid domains in monolayers}

\author{E. Velasco}
\email{enrique.velasco@uam.es}
\affiliation{Departamento de F\'{\i}sica Te\'orica de la Materia Condensada,
Universidad Aut\'onoma de Madrid,
E-28049, Madrid, Spain}

\author{L. Mederos}
\email{lmederos@icmm.csic.es}
\affiliation{Instituto de Ciencia de Materiales de Madrid, Consejo Superior de Investigaciones
Cient\'{\i}ficas, C/Sor Juana In\'es de la Cruz, 3, E-28049, Madrid, Spain}
\date{\today}

\begin{abstract}

We formulate a simple effective model to describe molecular interactions in a lipid monolayer. The model represents lipid molecules in terms of two-dimensional anisotropic particles on the plane of the monolayer.  These particles interact through forces that are believed to be relevant for the understanding of fundamental properties of the monolayer: van der Waals interactions originating from lipid chain interaction, and dipolar forces between the dipole groups of the molecular heads. Thermodynamic and phase behaviour properties of the model are explored using density-functional theory. Interfacial properties, such as the line tension and the structure of the region between ordered and disordered coexisting regions, are also calculated. The line tension turns out to be highly anisotropic, mainly as a result of the lipid chain tilt, and to a lesser extent of dipolar interactions perpendicular to the monolayer.  The role of the two dipolar components, parallel and perpendicular to the monolayer, is assessed by comparing with computer simulation results for lipid monolayers.

\end{abstract}

\maketitle

\section*{Introduction}

Lipid monolayers have been intensely investigated in the last decades because of their
importance as paradigms for various interfacial problems of biological importance, in particular
for the lung surfactant system \cite{Perez-Gil,Casals}. 
Monolayers made of one- or multi-component lipid molecules 
at the air-liquid water interface are considered as a useful model for a lipid monolayer.
One of the more extensively analysed lipid monolayer consists of DPPC molecules and
mixtures with similar molecules of different saturation degree
in their aliphatic chains. Below some critical point, pure and mixed monolayers generally show
phase separation between a phase with disordered chains, LE or Liquid-Expanded (fluid) phase, and
a phase with ordered chains, LC or Liquid-Condensed (gel- or solid-like)
phase \cite{Kaganer}. 

The molecular structure in the LC phase consists of molecules with straight and tightly packed molecular
chains. Molecules show a high degree of positional order, compatible with a global
or local two-dimensional crystal or glassy state, although the precise molecular ordering is open to debate. 
In the LE phase, by contrast, not only 
the chains but also the molecular centres of mass are disordered.
Also, molecular chains in both phases are observed to be tilted with respect to the monolayer normal to some degree. The tilt
is believed to optimise chain contact and van der Waals interactions. In the 
case of the LC phase the optimal contact energy compensates
for the decrease of entropy associated with the positional molecular
ordering. The LC-LE phase transition can conceptually 
be regarded as a
classical first-order phase transition between a two-dimensional orientationally disordered
liquid and a two-dimensional ordered crystal.

Lipid molecules are amphiphilic in nature, i.e. they show polar and nonpolar characteristics
that explain their tendency to occupy the liquid interface. These characteristics may 
play slightly different roles in the two-dimensional phase transitions. The
non-polar part, through the condensation and ordering of the molecular chains, do clearly
play the most important role in the phase transition, but less is known about the role of the
polar heads. Experimental investigation of this problem is difficult and consequently
scarce. Some important questions about lipid domains at the phase transition are still
open, for example, the intrincate domains shapes, the domain structure, and the stability and growth
kinetics of domains at coexistence. The very role played by the polar and nonpolar interactions
on the above properties is uncertain. 

From the theoretical side progress has been slow. To date theoretical models have been formulated
mostly at the mesoscopic level, and incorporate polar and nonpolar interactions between 
molecules more or less
implicitely \cite{domains1,domains2,Aurora}. 
Models have focused on the understanding of domain shape and domain-shape transitions, 
taking thermodynamic coexistence as given. But thermodynamically consistent {\it microscopic} 
models that start from an interaction potential energy are scarce, and existing models are 
formulated on lattices and do not include dipolar interactions
\cite{Mouritsen1,Mouritsen2}. More complete models, in the tradition of classical liquid-state
theory, are more powerful in that, not only bulk thermodynamics (and therefore phase
transitions), but also interface thermodynamics and structural information, can be calculated.
This information may be very useful as an input to mesoscopic models or to understand computer
simulations.

In the present paper we formulate a simple microscopic model based on interacting 
two-dimensional effective anisotropic particles. The model is inspired by recent
computer simulation results that use atomistic force fields \cite{us,Javanainen}. 
The model includes van der Waals interactions between lipid chains, and dipolar forces with
perpendicular and parallel components with respect to the monolayer, associated to polar interactions
between lipid head groups. Using density-functional theory we obtain phase diagrams for different
values of interaction parameters. In particular, we study their effect on the density
gap between coexisting domains. Also, the theory allows for the study of the microscopic structure at the 
interface separating the two coexisting domains and the molecular orientation at the interface. 
The line tension turns out to be strongly anisotropic with respect to this orientation. 

Our results have implications at various levels. First,
we conclude that the model correctly reproduces the molecular orientation at the LC domain 
boundaries if the inplane component of molecular dipoles is absent. Therefore, our results support the
concept that this dipolar component should play no role in determining the structure. This conclusion 
is in agreement with the atomistic simulation 
results \cite{Javanainen}. Second, the dipolar component perpendicular to the monolayer does not essentially perturb
molecular orientation at the boundary, although it substantially lowers the line tension. Also, anisotropicity
of the line tension with respect to molecular orientation at the boundary, already present in the absence of any
dipolar interaction, is marginally reinforced by the presence of a perpendicular dipole. The anisotropic nature of the
line tension is an essential feature of the model, and we argue that it should be incorporated in mesoscopic
models for domain shape.

\section*{Theoretical section}

\begin{figure}[h]
\begin{center}
\includegraphics[width=0.75\linewidth,angle=0]{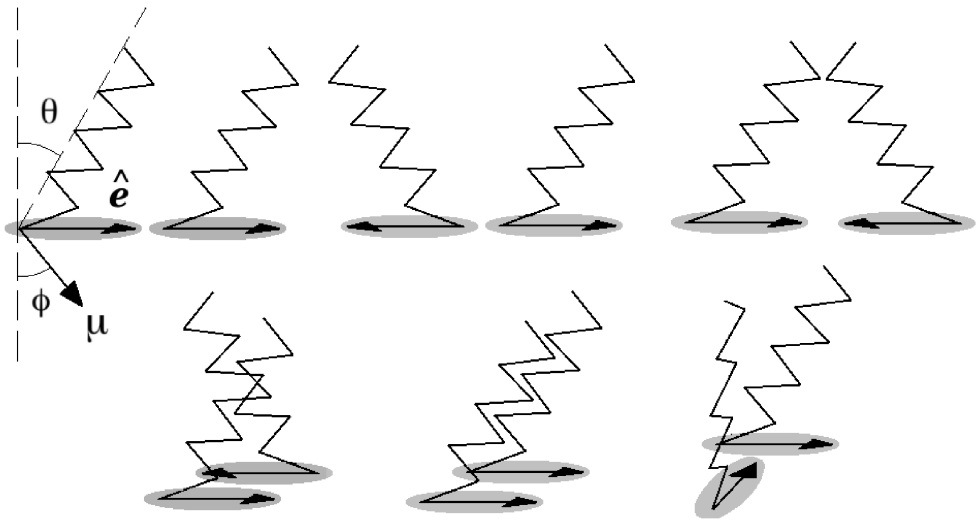}
\caption{\label{fig1} Schematic of representative 
effective-particle configurations and corresponding
lipid chain orientations. Details given in the text.}
\end{center}
\end{figure}

Some features of our two-dimensional effective model for a lipid monolayer are inspired by the recent
simulation study \cite{us} of an atomistic model for a DPPC monolayer. In this reference
the density of interaction units of a molecule, projected on the plane of the monolayer,
was studied separately for domains of the LC and LE phases. A histogram of 
projected molecular aspect ratios was obtained, where the aspect ratio was defined from the two
gyration radii of the projected interaction units of the molecules. Despite the different chain orderings 
in the LC and LE domains, the respective histograms result in mean aspect ratios very close to 2.5
in both phases. 
This means that the average shape of a lipid molecule, projected on the plane of the monolayer,
is close to that of an ellipse with the latter aspect ratio, Fig. \ref{fig1}.

Based on these results, in the present model 
we assume that our two-dimensional effective particles consist of identical rectangular particles with
an aspect ratio of 3. The reason why a rectangular, rather than elliptical, shape is chosen, and for
the value of aspect ratio adopted, will become clear later on. The elongated projected shape of 
lipid molecules reflects the molecular geometry
and also the tendency of molecules to be slightly tilted with respect to the monolayer normal in both
LC and LE phases. As regards the interaction between two molecules, the anisotropic shape captures the fact 
that side-to-side configurations have a shorter overlap distance than end-to-end configurations.
Therefore, the effective particle interaction contains a purely repulsive term represented by
a two-dimensional hard-rectangle (HR) potential for particles with length $L$ and width $\sigma_0$,
with $\kappa=L/\sigma_0=3$ the aspect ratio.

Fig. \ref{fig1} shows the most representative effective-particle configurations. 
Let us associate a
two-dimensional unit vector $\hat{\bm e}_i$ with the $i$th molecule. This vector is on the 
plane of the monolayer and points along the projection of the lipid tail. Therefore, it is parallel
to the long axis of the effective two-dimensional particle. 
If $\hat{\bm r}$ is the unit vector joining particle 1 with 2, we define
$a_i=\hat{\bm r}\cdot\hat{\bm e}_i$ and $b=\hat{\bm e}_1\cdot\hat{\bm e}_2$.
Then the three end-to-end configurations have $(b,a_1,a_2)=(+1,+1,+1)$, 
$(-1,\pm 1,\mp 1)$, while for the two side-to-side configurations
$(b,a_1,a_2)=(+1,0,0)$ and $(-1,0,0)$. With the chain molecular structure in mind, it is clear that,
as far as van der Waals interactions are concerned, the first two end-to-end configurations, 
$\rightarrow\rightarrow$ and $\rightarrow\leftarrow$,
cannot have the same energy (note that the configurations $\rightarrow\leftarrow$ and
$\leftarrow\rightarrow$ are taken as equivalent).
Also, the two side-to-side configurations, $\uparrow\uparrow$ and $\uparrow\downarrow$, 
cannot have the same energy either.
These two configurations contain the most important relative difference 
energetically: in the case $b=+1$ lipid chains are in complete contact, whereas in the 
second $b=-1$ they interact much less. 

Thus, to the HR potential, an anisotropic attractive contribution between the effective particles is
added. This contribution is intended to represent the van der Waals attraction between the lipid chains,
and should be sensitive to the sign of $b$.
The model for the attractive interaction will be a modified Gay-Berne (GB) potential
\cite{GB}. The GB potential
is an anisotropic Lennard-Jones-like potential which has been extensively studied in three-dimensional fluids
to account for interactions between anisotropic, mesogenic particles. However, its use in two-dimensional systems 
is scarce, probably because it predicts a continuous, rather than first order, phase transition between
isotropic and nematic fluids. Since the original GB model presents head-to-tail symmetry, it cannot
discriminate between the two side-to-side (or end-to-end) configurations, and a modification is required
to introduce a splitting between the configurations.

In the original GB model the potential energy for two
particles with relative centre-of-mass vector ${\bm r}={\bm r}_2-{\bm r}_1$ and 
orientations $\hat{\bm e}_1$ and $\hat{\bm e}_2$ is
\begin{eqnarray}
&&\Phi_{\rm GB}({\bm r},\hat{\bm e}_1,\hat{\bm e}_2)=
4\epsilon\left({\bm r},\hat{\bm e}_1,\hat{\bm e}_2,\right)
f\left({\bm r},\hat{\bm e}_1,\hat{\bm e}_2\right)
\label{eq1}
\end{eqnarray}
with
\begin{eqnarray}
\epsilon\left({\bm r},\hat{\bm e}_1,\hat{\bm e}_2,\right)
=\epsilon_{\rm GB}\epsilon_1^{\nu}\left(\hat{\bm e}_1,\hat{\bm e}_2\right)
\epsilon_2^{\mu}\left(\hat{\bm r},\hat{\bm e}_1,\hat{\bm e}_2\right)
\end{eqnarray}
and
\begin{eqnarray}
f\left({\bm r},\hat{\bm e}_1,\hat{\bm e}_2\right)=\left(\frac{\sigma_0}{r-\sigma\left(\hat{\bm r},\hat{\bm e}_1,\hat{\bm e}_2\right)+\sigma_0}\right)^{12}-
\left(\frac{\sigma_0}{r-\sigma\left(\hat{\bm r},\hat{\bm e}_1,\hat{\bm e}_2\right)+\sigma_0}\right)^{6}.
\end{eqnarray}
$\sigma\left(\hat{\bm r},\hat{\bm e}_1,\hat{\bm e}_2\right)$ is a contact distance that depends on
the relative particle angle $\hat{\bm r}$ and the two particle orientations
$\hat{\bm e}_1,\hat{\bm e}_2$. $\epsilon_{\rm GB}$ is an energy scale, while $\mu=2$ and $\nu=1$
are exponents. $\sigma_0$ is made to coincide with the particle width. 
In the standard Gay-Berne model the contact distance is given by the
Hard-Gaussian Overlap (HGO) model. The full expressions are:
\begin{eqnarray}
&&\sigma\left(\hat{\bm r},\hat{\bm e}_1,\hat{\bm e}_2\right)=
\sigma_{\rm HGO}\left(\hat{\bm r},\hat{\bm e}_1,\hat{\bm e}_2\right)=\sigma_0\left\{1-\frac{\chi}{2}\left[
\frac{\left(a_1+a_2\right)^2}
{1+\chi b}+
\frac{\left(a_1-a_2\right)^2}
{1-\chi b}\right]\right\}^{-1/2},\nonumber\\\nonumber\\
&&\epsilon_1\left(\hat{\bm e}_1,\hat{\bm e}_2\right)=\left\{1-\chi^{2}b^2\right\}^{-1/2},\nonumber\\\nonumber\\
&&\epsilon_2\left(\hat{\bm r},\hat{\bm e}_1,\hat{\bm e}_2\right)=1-\frac{\chi'}{2}\left[
\frac{\left(a_1+a_2\right)^2}{1+\chi'b}
+\frac{\left(a_1-a_2\right)^2}
{1-\chi'b}\right],
\label{GB_fun}
\end{eqnarray}
where $\sigma_{\rm HGO}\left(\hat{\bm r},\hat{\bm e}_1,\hat{\bm e}_2\right)$ is the contact distance of the HGO model (which closely approximates that of hard ellipses).

In our modified model, we introduce two variations to the standard GB model, Eqns. (\ref{GB_fun}). The first is
forced by the fact that the HGO model for the hard core does not produce a first-order phase transition
between the isotropic and nematic phases. As discussed, this is a necessary requirement for the model in our
application to coexisting domains in the monolayer (note that the presence of an attractive contribution in
the GB model does not modify this scenario). To correct this unwanted feature,
we have modified the hard core and adopted a HR shape. This model is known to exhibit a
first-order transition from the isotropic phase to the uniaxial nematic phase in a range of aspect ratios
\cite{Yuri}. In particular,
for an aspect ratio $\kappa=3$, the phase transition is of first order, and gives rise to a density gap
with a sufficiently wide coexistence region. The modification involves substituting the contact distance in
(\ref{GB_fun}) by that of the HR model, 
$\sigma\left(\hat{\bm r},\hat{\bm e}_1,\hat{\bm e}_2\right)=
\sigma_{\rm HR}\left(\hat{\bm r},\hat{\bm e}_1,\hat{\bm e}_2\right)$.

\begin{figure}[h]
\begin{center}
\includegraphics[width=0.35\linewidth,angle=-90]{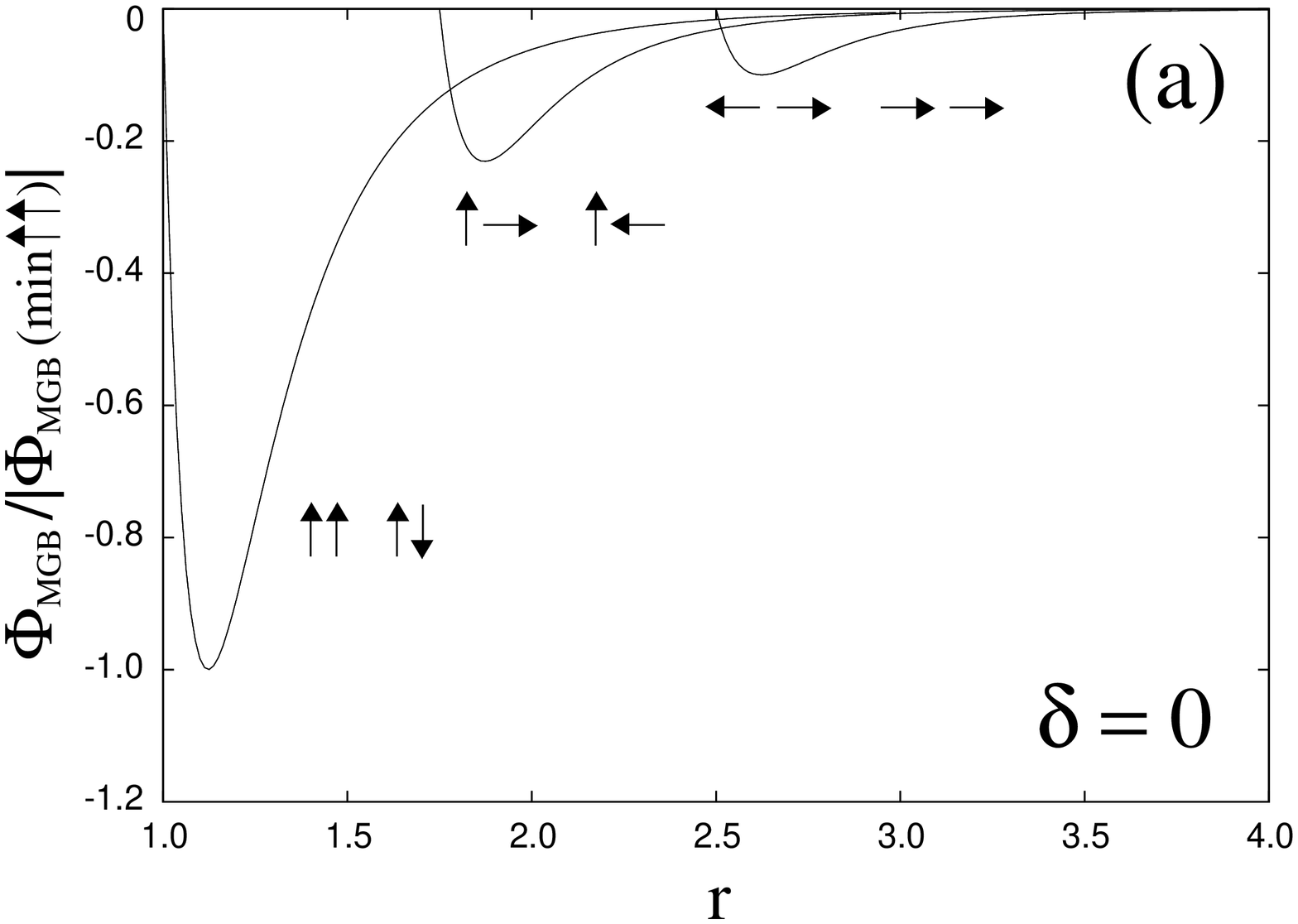}
\includegraphics[width=0.35\linewidth,angle=-90]{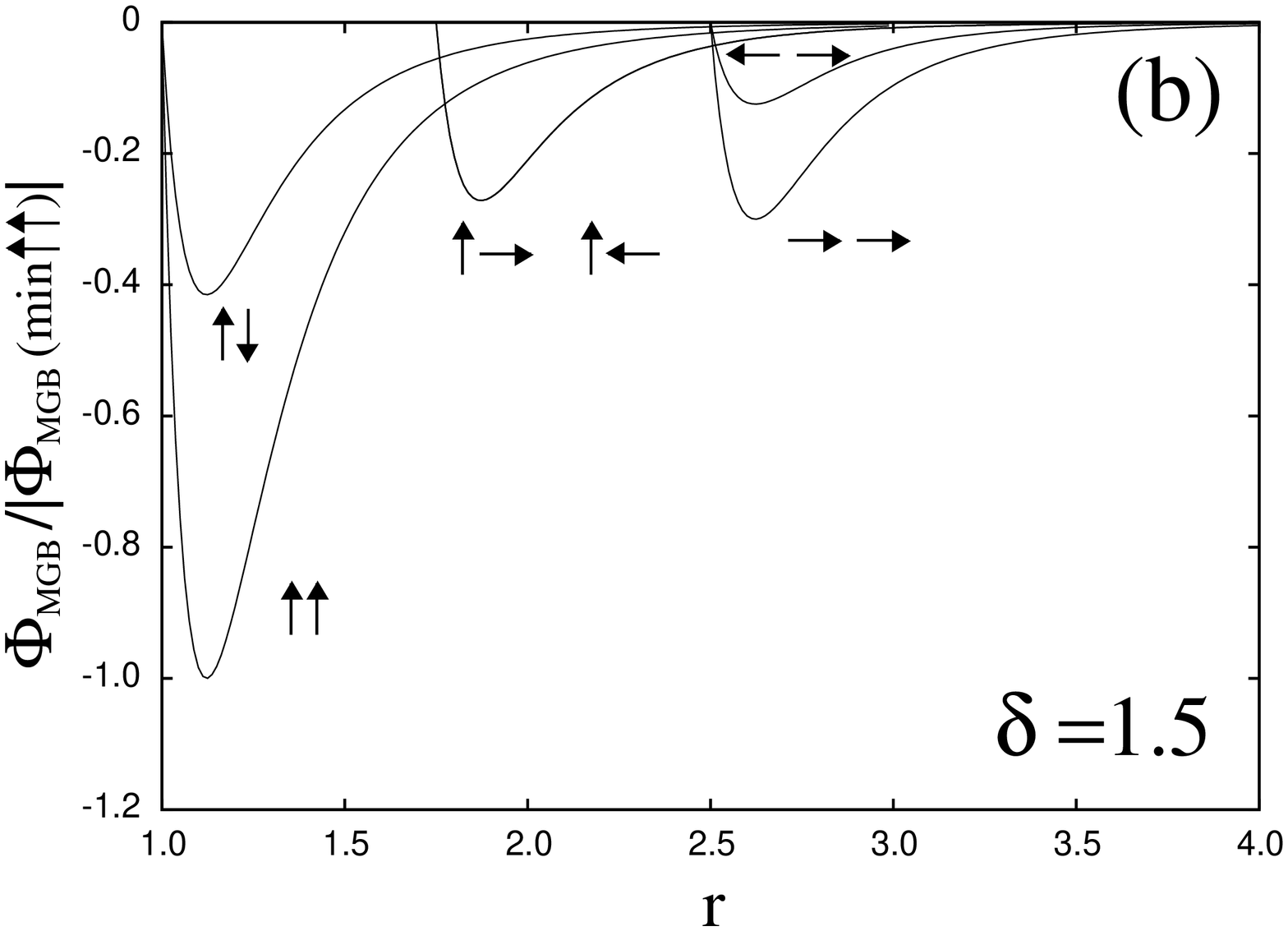}
\caption{\label{fig2} Modified Gay-Berne potential for various relative 
configurations of two molecules. $\hat{\bm e}$ vectors for each
configuration are indicated as arrows. Values of the parameters are
$\kappa=3$, $\kappa'=0.3$, $\mu=2$ and $\nu=1$. (a) $\delta =0$. (b) $\delta=1.5$.}
\end{center}
\end{figure}

The second modification involves splitting the energy of parallel $b=+1$ and antiparallel 
$b=-1$ configurations, as discussed above. The simplest way to implement this
feature is by modifying the $\epsilon_1$ function to
\begin{eqnarray}
\epsilon_1\left(\hat{\bm e}_1,\hat{\bm e}_2\right)=\left\{1-\frac{1}{2}\chi^{2}(1+\delta b)\right\}^{-1/2},
\label{MGB}
\end{eqnarray}
where $\delta$ is an asymmetry parameter that quantifies the energy splitting.
The resulting modified potential (\ref{eq1}) with (\ref{MGB}) instead of (\ref{GB_fun}) for the energy
function $\epsilon_1$ will be called {\it Modified Gay-Berne} (MGB) potential, 
$\Phi_{\rm MGB}({\bm r},\hat{\bm e}_1,\hat{\bm e}_2)$.
Note that $\delta=0$ does not reduce (\ref{MGB}) to the standard GB model. 

Fig. \ref{fig2} shows the MGB potential for various selected relative configurations of two particles:
the two side-to-side configurations $\uparrow\uparrow$ and $\uparrow\downarrow$, the two T configurations
$\uparrow\leftarrow$ and $\uparrow\rightarrow$, and the two side-to-side configurations
$\rightarrow\rightarrow$ and $\rightarrow\leftarrow$. The parameter $\kappa'$ describes the relative
energy between parallel and T configurations in the original GB potential, and is set to $0.3$.
In the figure results for two different values of the $\delta$ are shown:
$\delta=0$, panel (a), and $\delta=1.5$, panel (b).
We see how the two side-to-side configurations are degenerate for $\delta=0$ but split into
two distinct levels for $\delta>0$. The end-to-end configurations are also affected by
$\delta$, but not the T configurations, which remain degenerate.


Therefore our model represents lipid molecules as two-dimensional elongated objects 
(on the plane of the monolayer), interacting through anisotropic van der Waals forces.
Note that, in our model, interactions between the effective particles do not depend on the 
phase these particles belong to. This means that interactions are not sensitive 
to the chain disorder of molecules in the LE phase, as opposed to the perfect chain order 
in the LC phase. 
This shortcoming of the model could be overcome by considering an additional
particle variable accounting for the mean number of defects along the lipid chains. The
variable would be coupled to the other variables through modification of the 
energy prefactors in (\ref{GB_fun}).
However, we believe that the impact of assuming that interactions do not depend
on chain ordering is negligible in view of the
fact that our results for the line tension are in reasonable agreement with experiment, as
commented below.

In addition to the van der Waals interaction, an embedded linear dipole pointing along some direction
(not necessarily oriented in the plane) in included. This dipole represents 
the joint electrostatic charges of the neutral head group of the lipid molecule.
The pair potential energy is written as a sum of two contributions,
the modified Gay-Berne term and the dipole energy:
\begin{eqnarray}
\Phi({\bm r}_{12},\hat{\bm e}_1,\hat{\bm e}_2)=
\Phi_{\rm MGB}({\bm r},\hat{\bm e}_1,\hat{\bm e}_2)+
\Phi_{\rm dip}({\bm r},\hat{\bm e}_1,\hat{\bm e}_2).
\end{eqnarray}
The dipolar energy term is
\begin{eqnarray}
\Phi_{\rm dip}(\hat{\bm r},\hat{\bm e}_1,\hat{\bm e}_2)&=&\frac{\mu_{\perp}^2}{\epsilon_0\epsilon_{\parallel}r^3}+
\frac{\mu_{\parallel}^2}{\epsilon_0\epsilon_{\perp}r^3}
\left[\hat{\bm e}_1\cdot\hat{\bm e}_2-3\left(\hat{\bm e}_1\cdot\hat{\bm r}\right)
\left(\hat{\bm e}_2\cdot\hat{\bm r}\right)\right]=
\frac{\mu_{\perp}^2}{\epsilon_0\epsilon_{\parallel}r^3}+\frac{\mu_{\parallel}^2}{\epsilon_0\epsilon_{\perp}r^3}
\left(b-3a_1a_2\right).
\label{dipolar}
\end{eqnarray}
The dipole may have components normal and parallel to the monolayer,
respectively $\mu_{\perp}=\mu\cos{\theta}$ and
$\mu_{\parallel}=\mu\sin{\theta}$, where $\theta$ is the polar angle of the dipole moment. 
In this dipolar model we are assuming
that the dielectric constants in the two directions may be different. 
Atomistic simulations indicate that the angle $\theta$ in both LC and LE domains is
very close to $22^{\circ}$. As regards the parallel component, it is assumed to be aligned 
along the long axis of the effective projected particle, $\hat{\bm e}$. 
This assumption is at variance with the results of the latter simulation, which 
predicts a parallel component that librates in the azimuthal angle, essentially uncorrelated
with $\hat{\bm e}$. This result may be a feature of the force field used in the simulations
and should be confirmed. In our model this situation would correspond to setting the
in-plane component of the dipole $\mu_{\parallel}$ to zero.
Note that the model does not contemplate the situation where $\hat{\bm e}$
is not in the plane spanned by the dipole moment and the normal to the
monolayer.


With this model we attempt to describe a phase diagram involving two phases, one with orientational disorder
(equivalent to the LE phase of the lipid monolayer) and another with orientational order (akin to the LC phase).
Also, the effect of the dipolar strength on the coexistence gap can be studied, along with
the stability and shape of domains, and eventually some dynamical aspects of domain growth. Some of these issues are 
addressed in the present study. Others will be left for future work. 

As formulated, the model is expected to present a number of stable equilibrium phases. 
Among these are the isotropic phase, where molecules are both spatially and orientationally disordered, and 
the nematic phase, a fluid phase where molecules are oriented on average along some common direction, called the
director. The model also contains an exotic nematic phase, the tetratic, which is a fluid phase
with two equivalent directors. The tetratic can replace the standard nematic phase when the aspect ratio
of the particles is sufficiently low. Also the model exhibits a crystalline, fully ordered phase at high density.
The existence of this sequence of stable phases is based on the known properties of the hard core interaction
of the model, the HR model, which are well known. 
These and related models have been studied theoretically using mean-field theory \cite{Yuri,B3} and by means 
of simulation \cite{Stillinger}. In contrast, the effect of the addition of an anisotropic attractive interaction and 
a linear dipole has not been investigated. Although the nature of the stable phases is not expected to be modified
(based on similar models in two and three dimensions), the effect of the new ingredients will certainly be a profound 
one. In particular, we would like to obtain some trends as 
to the effect of the dipole orientation and strength. Initially we formulate a general model and explore
the consequences on phase behaviour.

The model is here analysed using classical density-functional
theory (DFT) and associated mean-field approximations to examine the phase behaviour.
At this point we neglect non-uniform spatial ordering since the introduction of this order in the theory
gives rise to an unnecessarily complicated numerical problem. Therefore, at the DFT level we restrict ourselves to a description of 
the uniform phases, i.e. isotropic and nematic (we choose the aspect ratio of the particles in such a way as to avoid
the stabilisation of tetratic ordering, which is certainly not observed in lipid monolayers). 
These phases will be identified respectively with the LE and LC phases. 

One crucial assumption of our approach concerns the identification of the phases. The isotropic
 phase of the model is identified with the Liquid Expanded (LE)
phase of the monolayer, while the nematic phase is meant to represent the Liquid Condensed (LC) phase. This may not be a realistic
representation, especially in the case of the LC phase, but at least the model has two essential ingredients: in the LC phase
the aliphatic chains of the molecules are rigid and oriented to optimise the van der Waals energy; and the components of the 
dipolar moment on the monolayer should be aligned and contribute to a global dipolar moment.

We now formulate a perturbation theory, taking the HR model as
reference system. We write the one-particle density,
Without loss of generality, as $\rho({\bm r},\hat{\bm e})=\rho({\bm r})f({\bm r},\hat{\bm e})$, 
where $\rho({\bm r})$ is the local number density and 
$f({\bm r},\hat{\bm e})$ the orientational distribution function. In the isotropic phase $f({\bm r},\hat{\bm e})=\frac{1}{2\pi}$,
since the distribution function has to be normalised:
\begin{eqnarray}
\int d\hat{\bm e}f({\bm r},\hat{\bm e})=\int_0^{2\varphi}d\varphi f({\bm r},\varphi)=1.
\end{eqnarray}
Now the potential energy is split into hard 
repulsive and attractive parts using a Barker-Henderson scheme.
Then we write a free-energy functional as a sum of ideal and excess part, with the latter containing
contributions from the MGB and dipolar terms:
\begin{eqnarray}
F[\rho]=F_{\rm id}[\rho]+F_{\rm HR}[\rho]+F_{\rm MGB}[\rho]+F_{\rm dip}[\rho]
\label{full_functional}
\end{eqnarray}
where
\begin{eqnarray}
\beta F_{\rm id}[\rho]&=&\int d{\bm r}\int d\hat{\bm e}\rho({\bm r},\hat{\bm e})\left\{\log{\left[
\rho({\bm r},\hat{\bm e})\Lambda^2\right]}-1\right\}\nonumber\\\nonumber\\
&=&\int d{\bm r}\rho({\bm r})\left\{\log{\left[\frac{\rho({\bm r})}{2\pi}\Lambda^2\right]}-1+
\int d\hat{\bm e}f({\bm r},\hat{\bm e})\log{\left[2\pi f({\bm r},\hat{\bm e}) \right]}\right\},
\label{Fid}
\end{eqnarray}
is the ideal free-energy contribution. $\Lambda$ is the thermal wavelength.
The hard-core contribution can be written as
\begin{eqnarray}
\beta F_{\rm HR}[\rho]&=&\psi_{\rm HR}(\rho_0)
\int d{\bm r}_1\int d\hat{\bm e}_1 \rho({\bm r}_1,\hat{\bm e}_1)
\int d{\bm r}_2\int d\hat{\bm e}_2
\rho({\bm r}_2,\hat{\bm e}_2)
v_{\rm exc}({\bm r}_2-{\bm r}_1,\hat{\bm e}_1,\hat{\bm e}_2),
\label{FHR}
\end{eqnarray}
where
\begin{eqnarray}
v_{\rm exc}({\bm r},\hat{\bm e}_1,\hat{\bm e}_2)=
\left\{\begin{array}{ll}\displaystyle 1,&r<\sigma_{\rm HC}(\hat{\bm r},\hat{\bm e}_1,\hat{\bm e}_2)\\
0,&r>\sigma_{\rm HC}(\hat{\bm r},\hat{\bm e}_1,\hat{\bm e}_2)
\end{array}\right\}=\Theta\left[\sigma_{\rm HC}(\hat{\bm r},\hat{\bm e}_1,\hat{\bm e}_2)-r\right]
\label{vexc}
\end{eqnarray}
is the overlap function, and $\rho_0$ is the mean density. 
The exact formulation of this functional depends on the hard-core 
model considered. In the case of HR, scaled-particle theory has been implemented 
in the past \cite{Yuri}, and this is the theory used here.  Also,
\begin{eqnarray}
F_{\rm MGB}[\rho]&=&\frac{1}{2}\int d{\bm r}_1\int d\hat{\bm e}_1\rho({\bm r}_1,\hat{\bm e}_1)
\int d{\bm r}_2\int d\hat{\bm e}_2
\rho({\bm r}_2,\hat{\bm e}_2)\nonumber\\
&&\hspace{2cm}\times\Theta\left[\left|{\bm r}_2-{\bm r}_1\right|-\sigma_{\rm HC}\left(\hat{\bm r}_{12},\hat{\bm e}_1,\hat{\bm e}_2\right)\right]
\Phi_{\rm MGB}({\bm r}_2-{\bm r}_1,\hat{\bm e}_1,\hat{\bm e}_2)
\label{FMGB}
\end{eqnarray}
is the attractive MGB contribution to the free energy. Finally
\begin{eqnarray}
F_{\rm dip}[\rho]&=&\frac{1}{2}\int d{\bm r}_1\int d\hat{\bm e}_1\rho({\bm r}_1,\hat{\bm e}_1)
\int d{\bm r}_2\int d\hat{\bm e}_2
\rho({\bm r}_2,\hat{\bm e}_2)\nonumber\\
&&\hspace{2cm}\times\Theta\left[\left|{\bm r}_2-{\bm r}_1\right|-\sigma_{\rm HR}\left(\hat{\bm r}_{12},\hat{\bm e}_1,\hat{\bm e}_2\right)\right]
\Phi_{\rm dip}({\bm r}_2-{\bm r}_1,\hat{\bm e}_1,\hat{\bm e}_2)
\label{Fdip}
\end{eqnarray}
is the dipolar contribution. In these expression we are assuming that correlations are given by a simple step function (i.e. only
the correlation hole is taken into account).

\begin{figure}[h]
\begin{center}
\includegraphics[width=0.55\linewidth,angle=0]{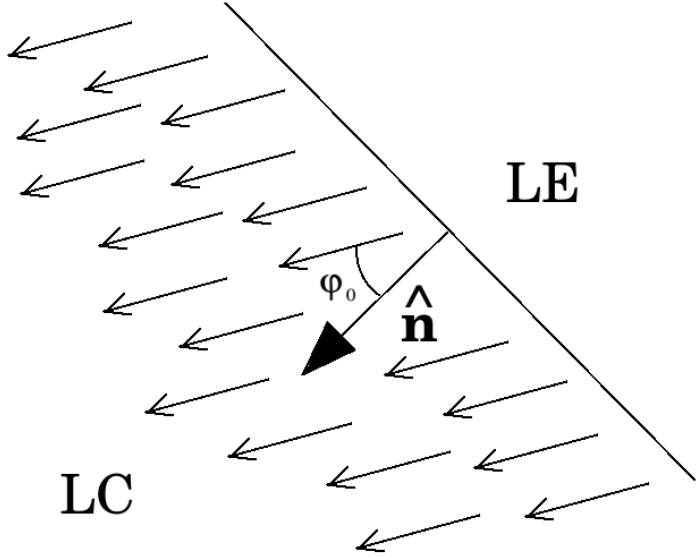}
\caption{\label{esquema} Definition of the angle $\varphi_0$ at the boundary between LE and LC
domains. Arrows represent the orientation of the local director, which is taken to be uniform
in our model. $\hat{\bm n}$ is the unit vector along the inward normal.}
\end{center}
\end{figure}

Now we give some details on how the theory is solved. We only sketch the
basic approximations and the numerical approach, considering the more general interfacial case;
the thermodynamics of the bulk phases can be obtained from the same interfacial approach,
using the corresponding bulk distributions. The interfacial structure is computed using a
variational method, where the density $\rho(x)$ and a set of orientational parameters
$\{\Lambda_n(x)\}$ are taken as variational functions.  In the context of the present
mean-field approach, we do not consider fluctuations of the boundary and instead assume a flat
boundary. The reference axis $x$ is taken along the boundary normal. This means that the variational
parameters, $\rho(x)$ and $\{\Lambda_n(x)\}$, will be functions of the $x$ coordinate only.
The orientational distribution function is then $f(x,\varphi)=f(\varphi;\{\Lambda_n(x)\})$,
where $\varphi$ is the angle between the long axis of a particle $\hat{\bm e}$ and the 
$x$ axis of the lab reference frame. The following parameterisation is used:
\begin{eqnarray}
f(x,\varphi)=\frac{\displaystyle e^{\displaystyle
\sum_{n=1}^{\infty}\Lambda_n(x)\cos{n\left(\varphi-\varphi_0\right)}}}
{\displaystyle\int_0^{2\pi} d\varphi' e^{\displaystyle
\sum_{n=1}^{\infty}\Lambda_n(x)\cos{n\varphi'}}}.
\label{Param_int}
\end{eqnarray}
In this expression we allow for the possibility that, in the nematic phase, the director is oriented
at an angle different from the normal direction. This effect is central to our discussion and is
taken into account through the angle $\varphi_0$ in (\ref{Param_int}), which implies a global rotation of the
director, see Fig. \ref{esquema}. The parameters $\{\Lambda_n(x)\}$ at each position $x$ will be a measure
of particle ordering about the direction dictated by $\varphi_0$. Note that in this work we do not
assume a spatial dependence of the angle $\varphi_0$, which means that the director is uniform
across the boundary with no deformation. This assumption is based on our choice of a flat
geometry for the boundary and on our inability to tackle the more general problem of boundary
deformations and fluctuations within the present formulation. We limit the number of
variational parameters to the first two parameters, $\Lambda_1(x)$ and $\Lambda_2(x)$,
which contain the symmetry of the polar and uniaxial nematic phases.
Once the equilibrium values of
$\Lambda_1(x)$ and $\Lambda_2(x)$ are obtained, the corresponding local order parameters
$s_1(x)$ and $s_2(x)$ in the proper frame (i.e. the frame associated with the nematic
director) can be obtained from 
\begin{eqnarray}
s_n(x)=\int_0^{2\pi} d\varphi f(x;{\varphi})\cos{n(\varphi-\varphi_0)}.
\end{eqnarray}
The minimisation is realised by discretising the spatial coordinates,
which turns the problem into a minimisation problem of a function of many variables, i.e. all the
values of $\rho$, $\Lambda_1$ and $\Lambda_2$ at the discrete points defined along the $x$ axis.
Due to the long-range nature of the dipolar interactions, many such points have be defined, meaning
that the computation box containing the interfacial boundary between the isotropic and nematic phases
will be very long in the direction normal to the boundary. The cutoff in the interactions is set to
$30\sigma_0$. We have used a box of length $100\sigma_0$ in the direction normal to the interface, 
with a grid size $\Delta x=0.1\sigma_0$ (the effective box length is in fact extended to $160\sigma_0$
to take into account the nonlocal interactions).

Using the property
of translational invariance in the direction of the flat boundary ($y$ axis), all integrals over
${\bm r}_1=(x_1,y_1)$ and ${\bm r}_2=(x_2,y_2)$ in Eqns. (\ref{FHR}), (\ref{FMGB}) and (\ref{Fdip})
can be written as integrals along $x$ coordinates only.
In the process, effective potentials $\tilde{v}_{\rm exc}(x,\varphi_1,\varphi_2)$,
$\tilde{\Phi}_{\rm MGB}(x,\varphi_1,\varphi_2)$ and
$\tilde{\Phi}_{\rm dip}(x,\varphi_1,\varphi_2)$, are defined, which can be computed in advance.
These effective potentials are already integrated in the $y$ coordinates, and incorporate the
correlation hole given by the Heaviside function. For example, in the case of the dipolar term, we have:
\begin{eqnarray}
F_{\rm dip}[\rho]&=&\frac{L}{2}\int_{-\infty}^{\infty} dx_1\rho(x_1)
\int_{-\infty}^{\infty} dx_2 \rho(x_2) \nonumber\\&\times&
\int_0^{2\pi}d\varphi_1 \int_0^{2\pi}d\varphi_2
f\left(\varphi_1;\Lambda_1(x_1),\Lambda_2(x_1)\right)
f\left(\varphi_2;\Lambda_1(x_2),\Lambda_2(x_2)\right)
\tilde{\Phi}_{\rm dip}(x_1-x_2,\varphi_1,\varphi_2).
\label{Feff_dip}
\end{eqnarray}
The factor $L$, which represents the length of the (flat) boundary, comes by invoking translational
invariance along the boundary, which is expressed by the presence of the factor $y_1-y_2$.
The calculation of integrals such as those in (\ref{Feff_dip}), which have the same structure in
the case of the other free-energy contributions, involves a serious numerical burden. We must bear
in mind that a minimisation process over many variables is imposed on the full free-energy functional,
and this process involves a very large number of free-energy evaluations.

Our strategy was to evaluate the double angular integrals in a single step before the minimisations.
For example, for the dipolar term, we evaluate
\begin{eqnarray}
&&V_{\rm dip}(\Lambda_1^{(1)},\Lambda_2^{(1)},\Lambda_1^{(2)},\Lambda_2^{(2)})\nonumber\\&&
\hspace{1cm}\equiv
\int_0^{2\pi}d\varphi_1 \int_0^{2\pi}d\varphi_2
f\left(\varphi_1;\Lambda_1(x_1),\Lambda_2(x_1)\right)
f\left(\varphi_2;\Lambda_1(x_2),\Lambda_2(x_2)\right)
\tilde{\Phi}_{\rm dip}(x_1-x_2,\varphi_1,\varphi_2),
\end{eqnarray}
and create a large table with four entries: the values of two parameters $\Lambda_1,\Lambda_2$
evaluated at the first particle (coordinates with subindex $1$) and the values of two parameters
$\Lambda_1,\Lambda_2$ evaluated at the second particle (coordinates with subindex $2$), i.e.
$\Lambda_1^{(1)}\equiv\Lambda_1(x_1)$, $\Lambda_2^{(1)}\equiv\Lambda_2(x_1)$,
$\Lambda_1^{(2)}\equiv\Lambda_1(x_2)$ and $\Lambda_2^{(2)}\equiv\Lambda_2(x_2)$. This table
is then interpolated for intermediate values of the parameters. The accuracy of this procedure is
reasonable. The same procedure
is applied to the other free-energy contributions. This strategy saves a lot of
computer time and simplifies the boundary calculations considerably.

The relevant free-energy function to minimise is the line tension, which is defined as
the excess grand potential per unit length, $\lambda=(\Omega-\Omega_0)/L$, 
where $\Omega=F-\mu N$ is the grand potential, $L$ the interface length,
$\mu$ the chemical potential at coexistence, $N$ the number of particles, and 
$\Omega_0$ is the bulk grand potential.
The line tension is minimised with respect to all the independent variables
defined on the discretised $x$ axis, $\rho(x)$,
$\Lambda_1(x)$ and $\Lambda_2(x)$, using a conjugate-gradient
method. In each case the angle $\varphi_0$ between the nematic director and the monolayer normal
is fixed at some value in the interval $[0^{\circ},180^{\circ}]$.
This process recovers the bulk results and bulk coexistence very accurately.
However, despite the numerical accuracy of our strategy, very small deviations exist for the
bulk properties at different values of $\varphi_0$. This is a problem since the computation box
is assumed to be coupled
to bulk isotropic and nematic phases at each side of the box, and definite coexistence values for
density and order parameters, consistent with the numericals of the interfacial problem,
have to be fixed as boundary conditions. Any minor difference in the boundary conditions will
be detrimental for the correct minimisation. The solution is to obtain the bulk coexistence
consistently for each value of $\varphi_0$, which ensures perfect matching and a smooth minimisation
process. 

For the bulk phases we assume the dependence
$\rho(x)=\rho_0$ and $f(x;\varphi)=f(\varphi)$ for the phase with the lowest symmetry, 
the nematic phase.
In the isotropic phase $f(\varphi)=1/2\pi$. In this case we numerically minimise the 
total Helmholtz free-energy functional,
using the same strategy as for the interface. Once the equilibrium (constant) values of the 
$\Lambda_1$ and $\Lambda_2$ parameters are obtained for the nematic phase, the 
order parameters and the equilibrium free energy can be evaluated. From this the
chemical potential and the pressure can be computed numerically, and the whole phase 
diagram obtained by applying the equal-pressure and equal-chemical potential
conditions at each temperature $T$.

\section*{Results and discussion}

For both bulk and interface three cases have been analysed:
zero dipole, dipole in the plane of the monolayer,
and dipole perpendicular to the monolayer. The asymmetry parameter $\delta$ is fixed to a
value $\delta=1.5$. This value gives an energy gap between parallel and antiparallel
configurations of $58\%$ (see Fig. \ref{fig2}). As discussed previously, we do not claim
this value to be representative of any realistic situation. We simply argue that $\delta$
reflects the asymmetry between
the two unequivalent configurations of two tilted molecular chains, and that larger tilt
angles may reasonably be associated with larger values of $\delta$. A proper connection
between the atomistic model and the effective two-dimensional model can be done, but is outside
the scope of this exploratory investigation. Larger values of $\delta$,
giving stronger energy anisotropies, on the other hand,
do not substantially change the results and the qualitative conclusions that can be drawn
from the model. In order to check this point, a value $\delta=1.75$, giving an energy gap of $69\%$,
was also explored. There is a technical problem with the value of the asymmetry parameter, since in
the model the value of $\delta$ cannot be chosen arbitrarily. For example, for clearly smaller values
the density gap of the isotropic-nematic transition becomes extremely small or even disappears, which
is not realistic for the present application. In practice we have checked that $1.5$ is slightly above
the limiting value below which a realistic density gap for the isotropic-nematic transition
cannot be obtained. What we mean by `realistic density gap' will be discussed below.

\begin{figure}[h]
\begin{center}
\includegraphics[width=0.35\linewidth,angle=-90]{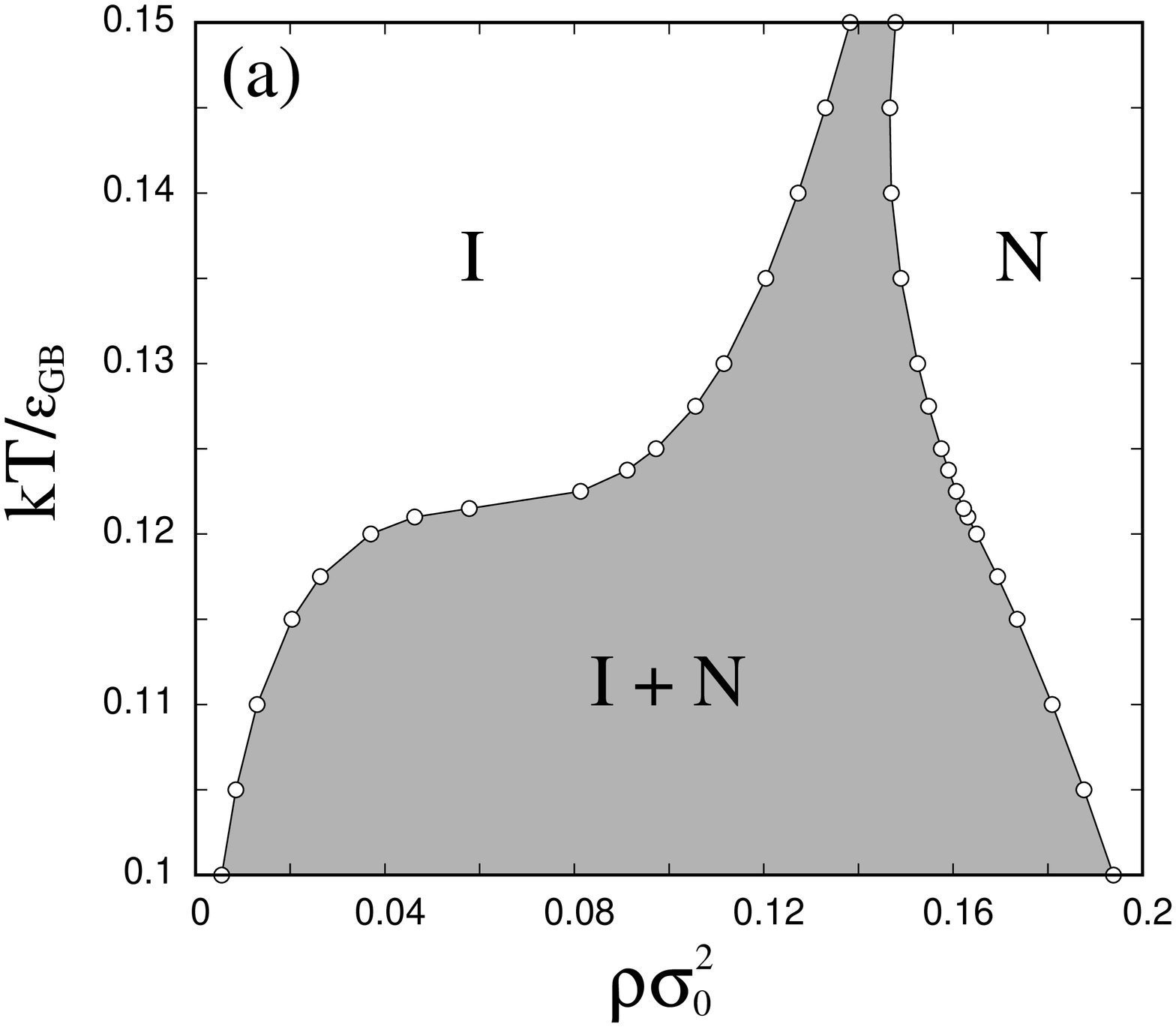}
\includegraphics[width=0.35\linewidth,angle=-90]{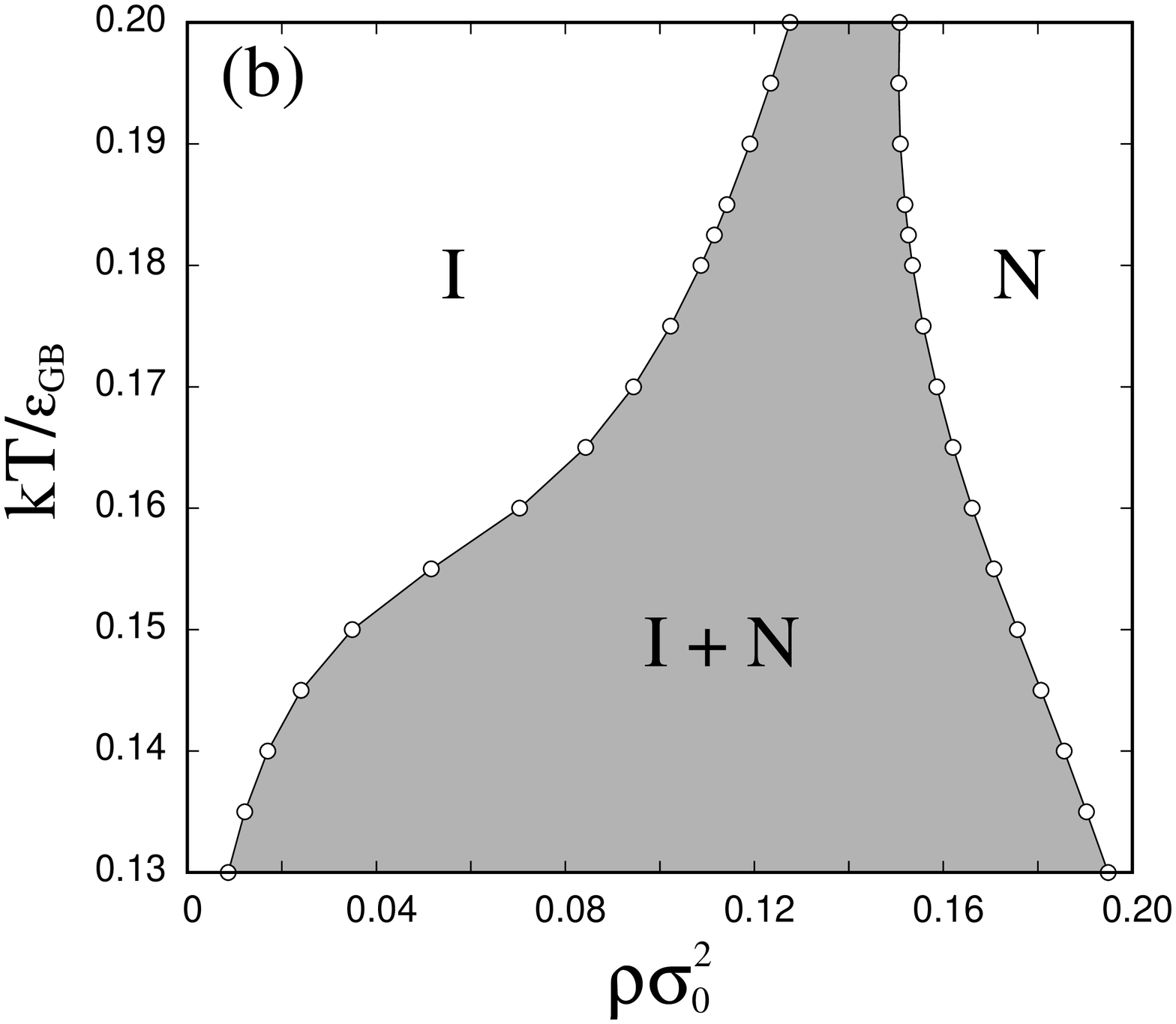}
\includegraphics[width=0.35\linewidth,angle=-90]{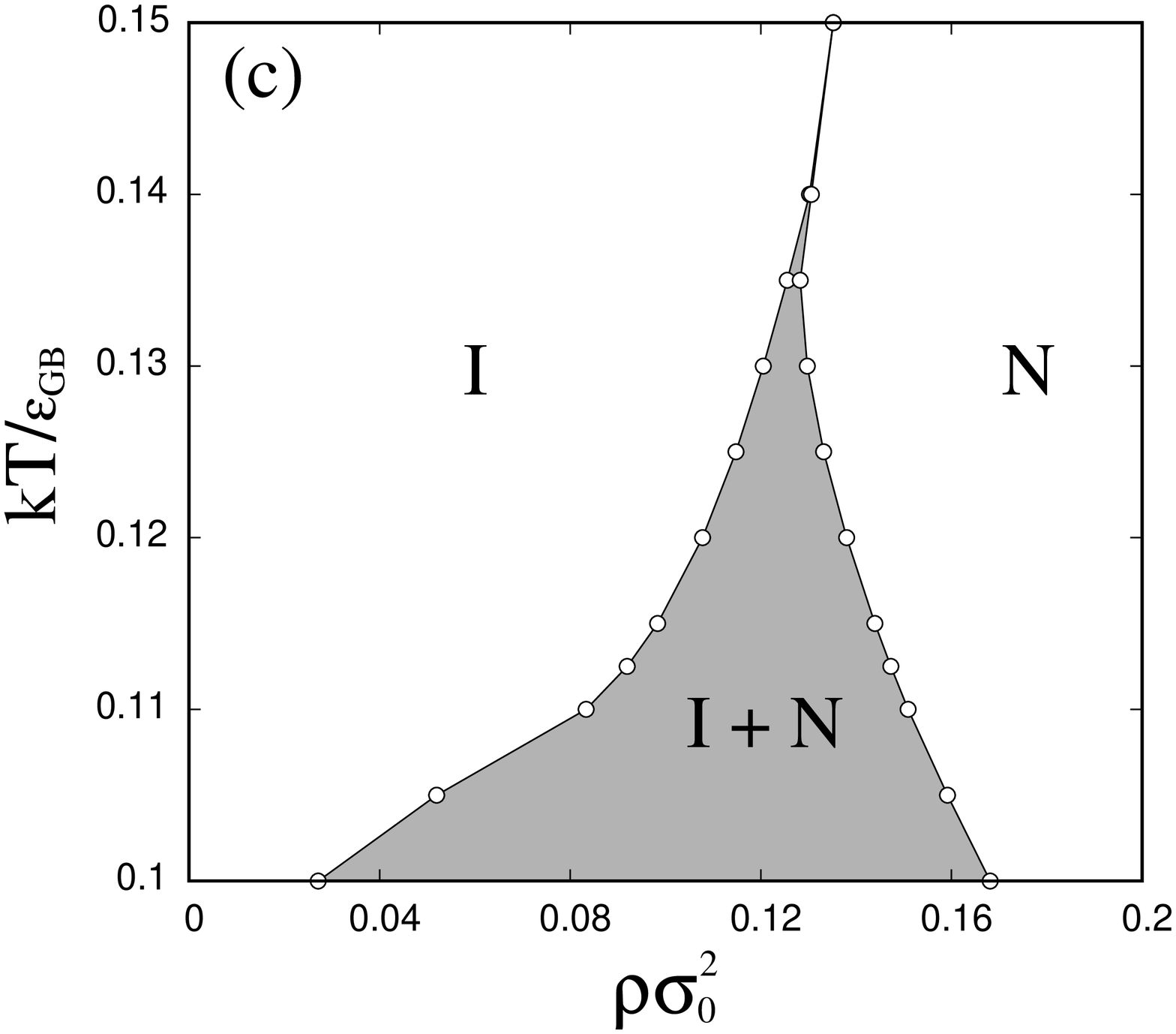}
\includegraphics[width=0.35\linewidth,angle=-90]{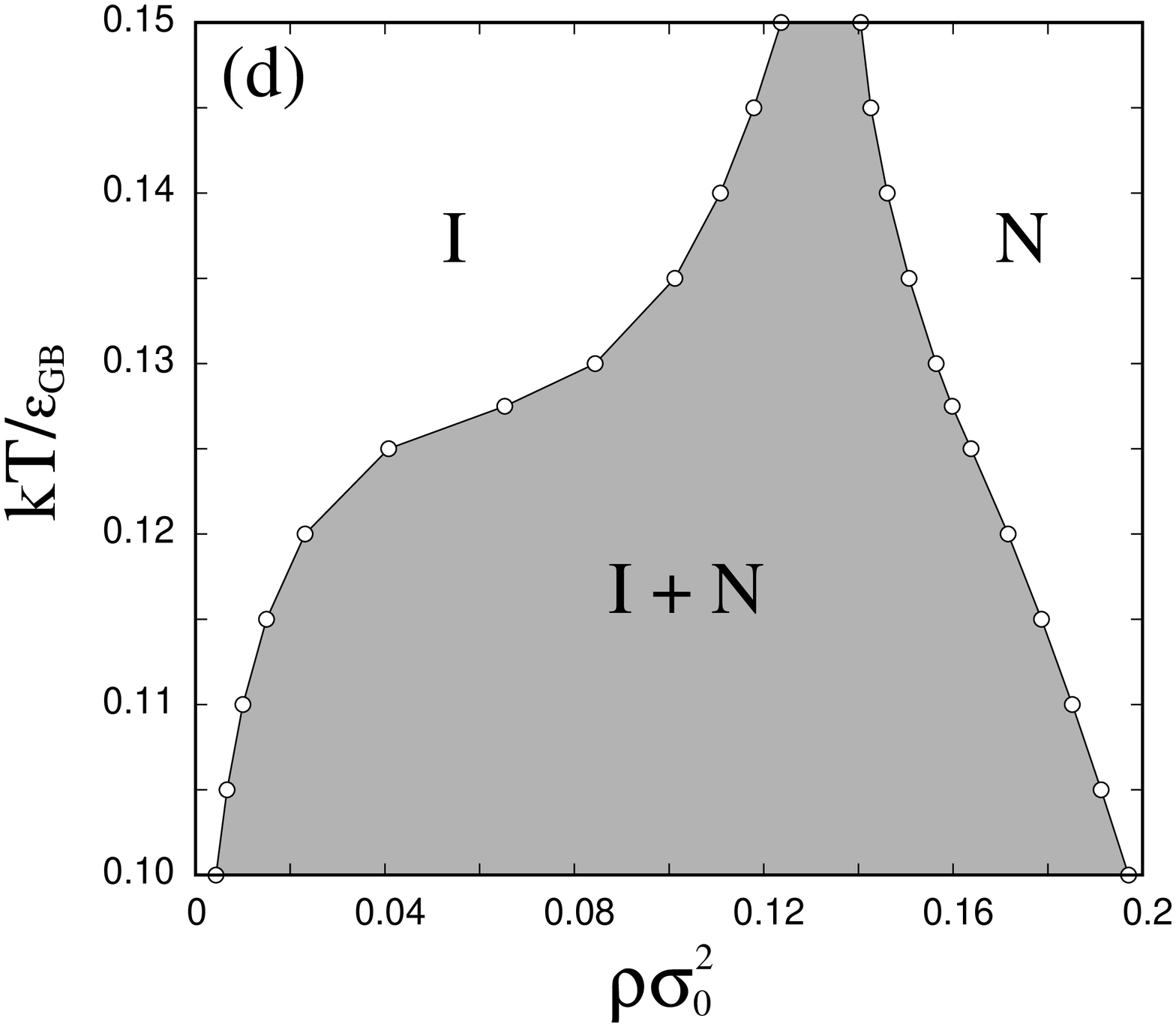}
\caption{\label{diags}Phase diagrams for different values of the parameters.
(a) $\delta=1.5$, $\mu^{*}_{\perp}=0$, $\mu^{*}_{\parallel}=0$.
(b) $\delta=1.75$, $\mu^{*}_{\perp}=0$, $\mu^{*}_{\parallel}=0$.
(c) $\delta=1.5$, $\mu^{*}_{\perp}=0.6$, $\mu^{*}_{\parallel}=0$. And (d)
$\delta=1.5$, $\mu^{*}_{\perp}=0$, $\mu^{*}_{\parallel}=0.6$.}
\end{center}
\end{figure}

Fig. \ref{diags} presents four phase diagrams in the temperature-density plane corresponding to the cases
mentioned above. All quantities are scaled with the appropriate parameters to give dimensionless
quantities, $\epsilon_{\rm GB}$ for energies and $\sigma_0$ for lengths. In the case of the dipolar
strengths we scale as $\mu_i^*=\mu_i/\sqrt{\epsilon_0\epsilon_i\epsilon_{\rm GB}\sigma_0^3}$,
with $i=\parallel,\perp$.
All phase diagrams present a common feature: the existence of a transition
between an isotropic (LE) phase and a nematic (LC) phase. In all cases the transition is of
first order and preempts the isotropic-vapour transition, which is metastable and below
the two isotropic-nematic binodal curves. The latter feature is not a limitation of the
model for the present study, which focusses on the phase coexistence between orientationally
ordered and disordered phases. Also in all cases the density gap between isotropic
and nematic phases rapidly decreases with
temperature and eventually closes, giving rise to a continuous phase transition at a
tricritical point. In real monolayers the density gap disappears at a critical point,
above which LC and LE regions can be continuously connected. The different symmetries of
the phases involved in our model and the absence of fluctuations in our mean-field treatment
precludes the existence of a critical point. Again this shortcoming is not an essential
point for our purposes. Note that, for even higher temperatures, the density gap has to
return to a finite value since the infinite-temperature limit of the model is the HR model,
which exhibits a first-order transition \cite{Yuri}.

We first compare panels (a) and (b) of Fig. \ref{diags}, both corresponding to zero dipole but
for different values of $\delta$. It is clear that an increasing anisotropy, producing a larger
energy gap, does not shift the isotropic-nematic transition in temperature, but gives rise to an
increased density gap. Similar conclusion are drawn from a comparison of panels (a) and (d),
corresponding to the same value of $\delta$ but to zero dipole and purely inplane dipole,
respectively. There exists an incipient isotropic-vapour transition, and the
inplane dipole slightly weakens the anisotropic interactions of the MGB model and
promotes the stability of the nematic phase. But more importantly, the inplane dipole
does not qualitatively affect the phase diagram.

By contrast, the case of a purely perpendicular dipole component, panel (c),
is qualitatively different. Condensation of the isotropic phase is clearly discouraged, since
the perpendicular dipole introduces a purely repulsive interaction. Also,
the effective anisotropic interactions promoting the ordering transition decrease, with
the result that the density gap at a given temperature is considerably reduced, and the
tricritical point (not visible in the scale of the other panels) moves to lower temperatures.

\begin{figure}[h]
\begin{center}
\includegraphics[width=0.35\linewidth,angle=-90]{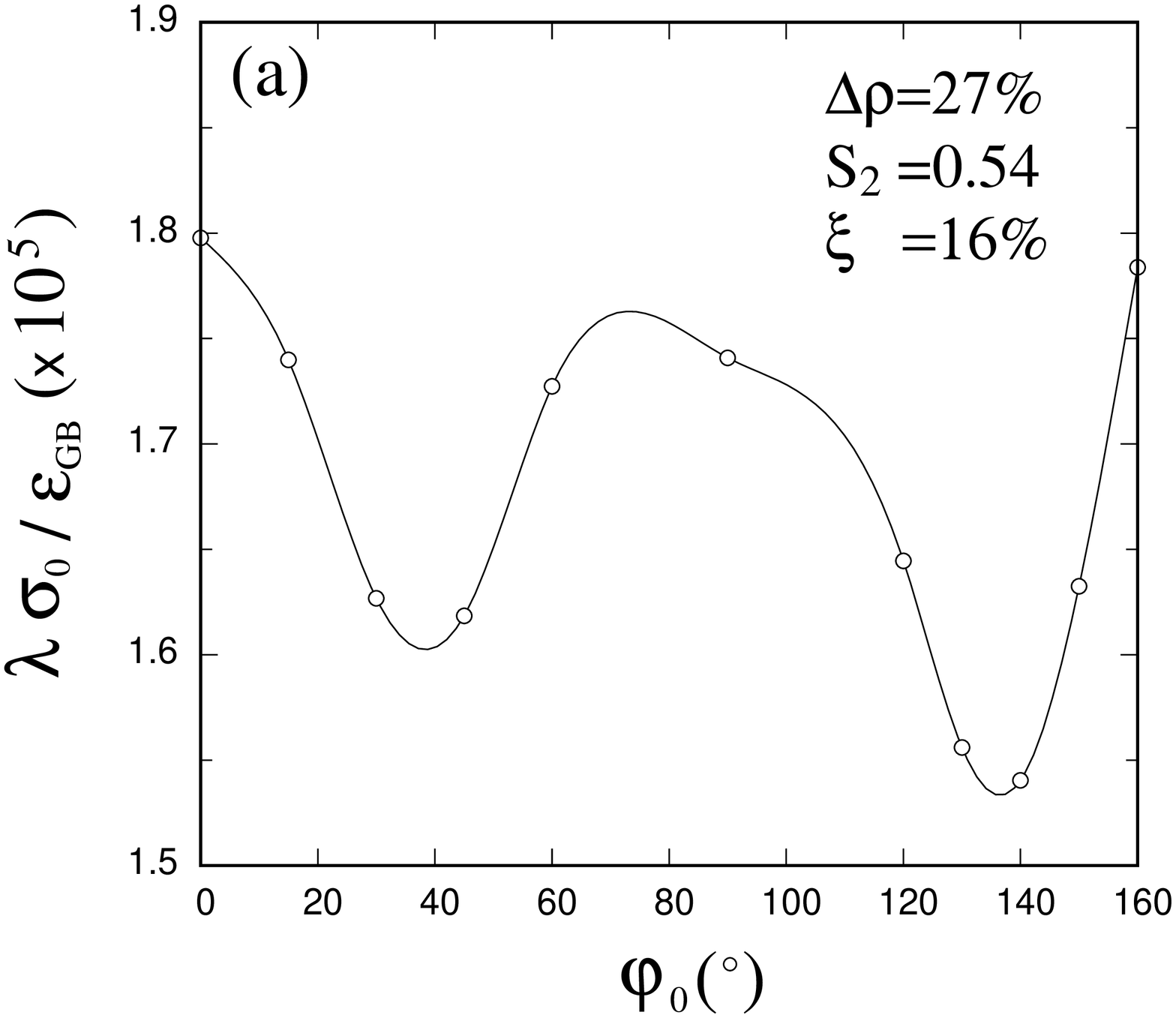}
\includegraphics[width=0.35\linewidth,angle=-90]{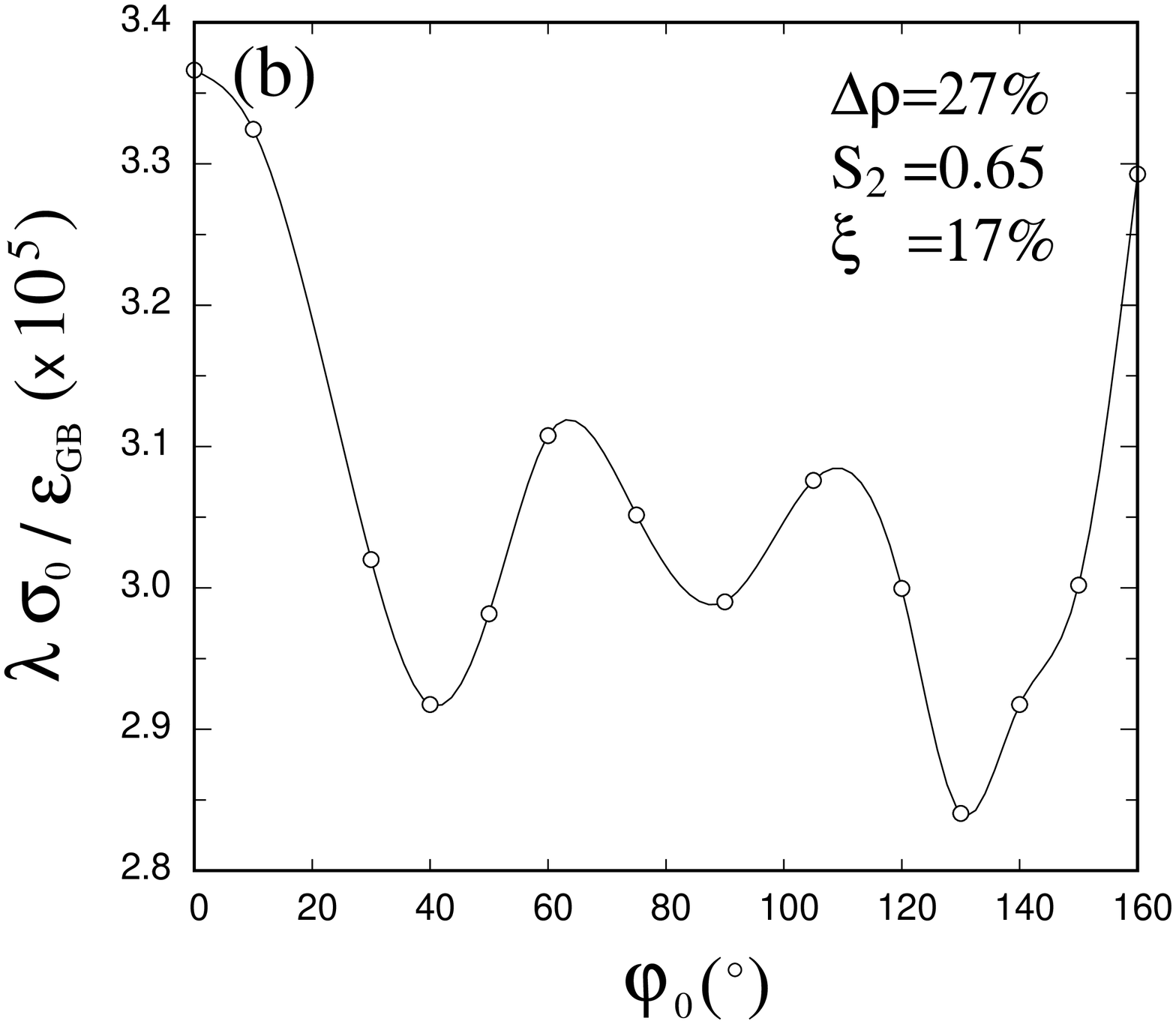}
\includegraphics[width=0.35\linewidth,angle=-90]{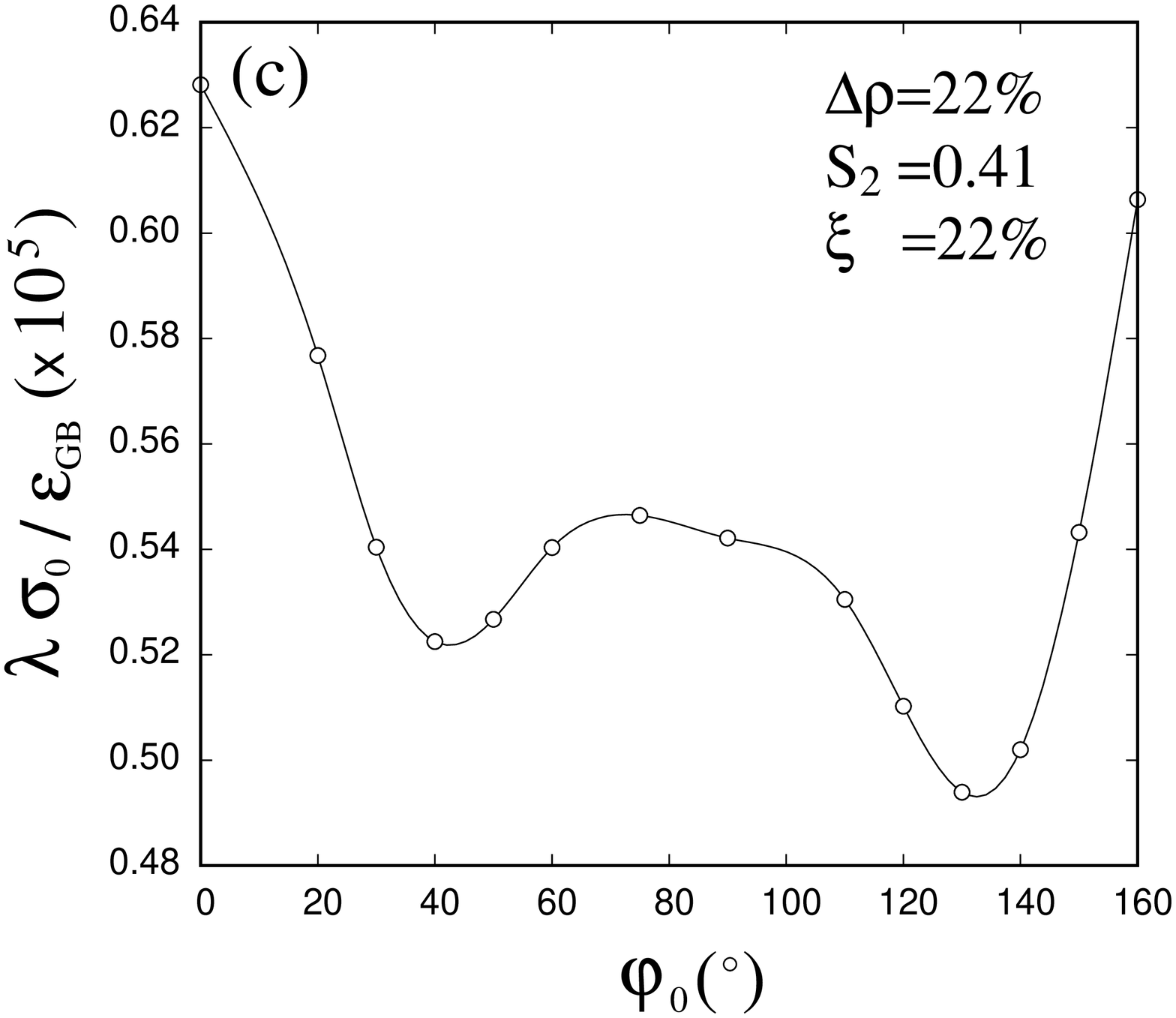}
\includegraphics[width=0.35\linewidth,angle=-90]{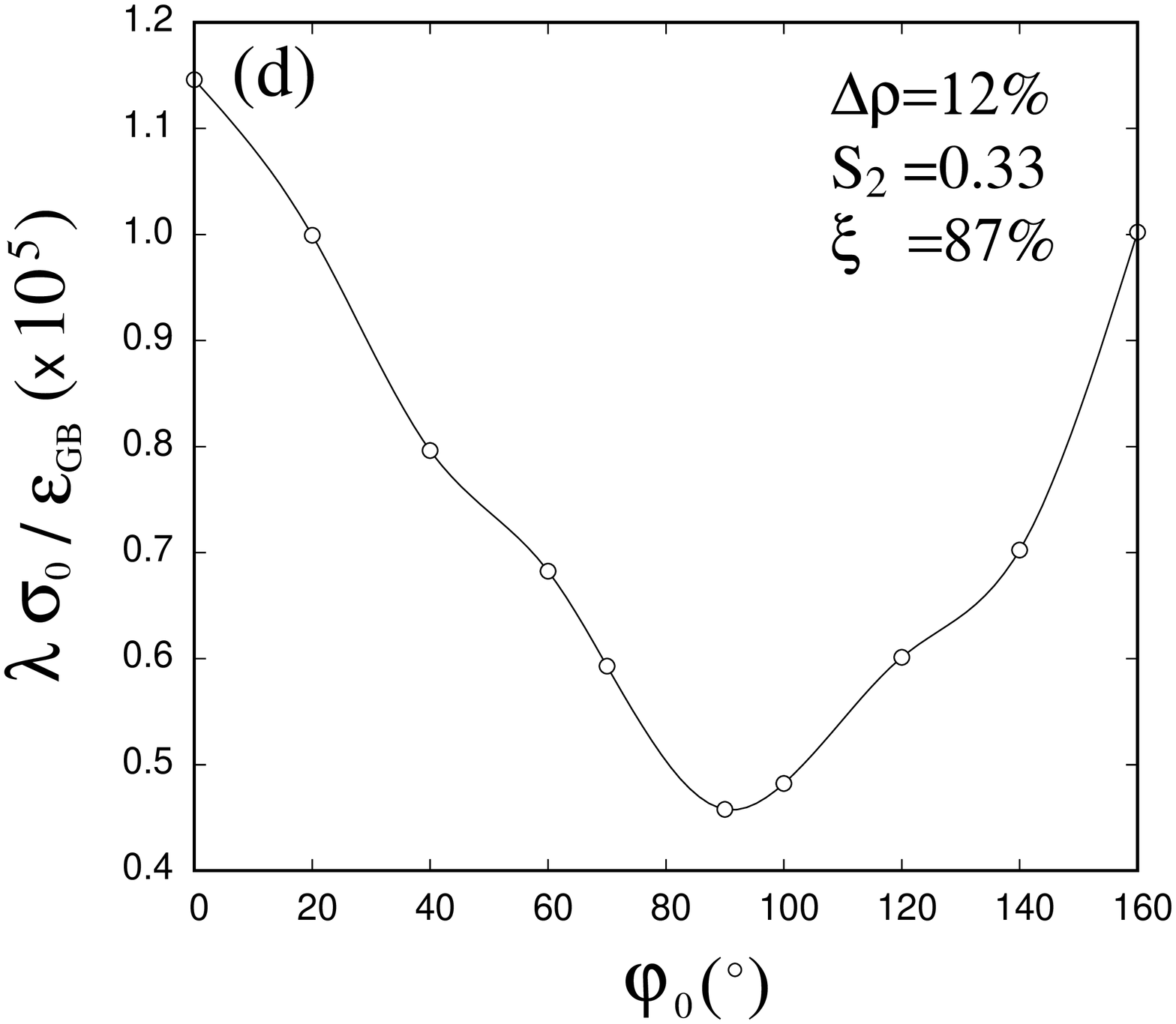}
\caption{\label{tension} Line tension as a function of the molecular orientation with respect to the
domain boundary normal. (a) $\delta=1.5$, $\mu^{*}_{\parallel}=0$, $\mu^{*}_{\perp}=0$
($kT/\epsilon_{\rm GB}=0.13$),
(b) $\delta=1.75$, $\mu^{*}_{\parallel}=0$, $\mu^{*}_{\perp}=0$
($kT/\epsilon_{\rm GB}=0.1825$). (c) $\delta=1.5$, $\mu^{*}_{\parallel}=0$, $\mu^{*}_{\perp}=0.6$ ($kT/\epsilon_{\rm GB}=0.12$).
(d) $\delta=1.5$, $\mu^{*}_{\parallel}=0.6$, $\mu^{*}_{\perp}=0$ ($kT/\epsilon_{\rm GB}=0.15$). Continuous line is a cubic spline
fitting intended only as a guide to the eye.}
\end{center}
\end{figure}

Fig. \ref{tension} presents the line tension $\lambda(\varphi_0)$ as a function of the angle
$\varphi_0$ between the average projection of the molecular chains in the LC domains (director)
and the normal to the
phase boundary $\hat{\bm n}$, see Fig. \ref{esquema}. The four panels correspond to the four cases shown in
Fig. \ref{diags}. As a reference, in each case the value of the temperature is chosen
such that the density gap is close or of the same order as that observed in the atomistic
simulations \cite{us}, which is approximately $27\%$ with respect to the density of the LC
phase. Therefore the temperatures are $kT/\epsilon_{\rm GB}=0.13$ in panel (a), $0.1825$ in
panel (b), $0.12$ in panel (c) and $0.15$ in panel (d), giving density gaps of $27\%$,
$27\%$, $22\%$ and $12\%$, respectively.

The first observation is that the line tension is considerably reduced when dipole interactions,
either perpendicular or parallel to the monolayer, are introduced. This reduction cannot be
explained alone by the density gap since their values are similar, see e.g. panels (a) 
and (c). Also, the
gap in the order parameters, e.g. the nematic order parameter $S_2$, at the transition, cannot
explain by itself the drop in line tension. The conclusion is that the dipolar interactions 
are responsible for the substantial reduction in $\lambda$. Experimental
values of $\lambda$ for lipid monolayers, extracted from domain size distributions
\cite{Israelachvili}, are in the range $1-10$ fN (although values measured using other techniques
may be higher by even two orders of magnitude depending on the system 
\cite{CWT,line_tension1,line_tension2}). Using $\sigma_0\sim 0.4$ nm (estimated
from experimental values of area per lipid in LC domains of DPPC monolayers) and 
$\epsilon_{\rm GB}\sim 50$ kT (estimated from atomistic force-field simulations \cite{us}, and
see also \cite{Libro_Israel}), we obtain values of $\lambda\sim 2-15$ fN, i. e. within the same order 
of magnitude as in \cite{Israelachvili}.

Interestingly, line tensions exhibit well-defined minima which correspond to equilibrium
oblique configurations at the boundary between LE and LC phases.
This is indicating the existence of an anisotropic line tension.
In cases where no dipole exists, panels (a) and (b), there is a global minimum at 
$\varphi_0\simeq 135^{\circ}$ and a local minimum at $\varphi_0\simeq 40^{\circ}$. These
two angles are close to being supplementary angles, which means that, in both
configurations, the long axes of the effective particles lie on the same straight line,
but the projected chains point in opposite directions. The most stable configuration
corresponds to chains pointing towards the bulk of LE regions at an angle of $135^{\circ}-90^{\circ}=45^{\circ}$
with respect to the boundary. Panel (c) corresponds to a model
where a dipole perpendicular to the monolayer is added. Clearly, the barrier between
minima is reduced, but the location of the minima does not change. This indicates that
a perpendicular dipole does not affect the anisotropy introduced by the MGB component.
Finally, in panel (d) a purely inplane dipole has been added. In this case the 
situation drastically changes, as only one minimum is visible at a angle of $\varphi_0=90^{\circ}$.
This corresponds to molecular projections parallel to the phase boundary.
In all cases the line tension is anisotropic, meaning that the interface pins the molecular
orientation, which is transmitted to the bulk via elastic forces, a situation similar to
the anchoring phenomena in liquid crystals.

Another interesting question is the degree of anisotropy of the line tension.
A possible measure of the anisotropy, $\xi$, is given by the amplitude of its variation in the
whole range of $\varphi_0$, $\Delta\lambda$, divided by the mean value $\lambda_0$, i.e.
$\xi=\Delta\lambda/\lambda_0$. In the cases shown in Fig. \ref{tension} the values
for the anisotropy are $\xi=0.16$, $0.22$ and $0.87$ in panels (a), (c) and (d), respectively.
Dipolar interactions parallel to the monolayer have a profound impact on the line-tension
anisotropy, whereas the perpendicular components also add some anisotropy over that of the
pure MGB potential which, at $\xi=16\%$, is already substantial. We are not aware of 
any experimental values for the anisotropy.

These results should be compared with available evidence on molecular structure and ordering.
An important source of information comes from atomistic computer
simulations. Two such simulations on DPPC monolayers have recently been published
\cite{Javanainen,us}. The two use the same force fields and the only difference is system
size. We focus on Ref. \cite{Javanainen}, in which
complete phase separation was probed along the coexistence region,
and pictures of representative molecular configurations were shown. Invariably, an angle
$\varphi_0$ at the domain boundaries different from $90^{\circ}$ can be seen in the
configurations. Direct visual estimates of the angle provide a value of $\sim 27^{\circ}$
(see panel 3, Fig. 2 of Ref. \cite{Javanainen}).

The above results imply that the inplane dipolar component cannot be large,
or at least cannot influence the structural properties of the domain. This is supported 
by the results of Ref. \cite{us} that the inplane
component of the dipoles show almost complete librational motion about the monolayer normal. 
This could imply that the inplane dipolar components of neighbouring molecules
interact weakly, because of strong screening \cite{Moy} or otherwise. 
Therefore, the only
relevant or effective dipolar component would be the perpendicular one.
The line tension is strongly affected by this component, which reduces its value. 
The angular anisotropy of the line tension, 
which is otherwise contained in the model with no dipole interactions,
increases moderately when perpendicular dipole forces are present. 
However, the main contribution of the anisotropy is due to the molecular tilt,
and the perpendicular dipoles only contribute to a lesser extent. 
The angular dependence of the line tension can be represented by
\begin{eqnarray}
\lambda(\varphi_0)=\lambda_0+\sum_{n=1}^{\infty}\lambda_n\cos{n\varphi_0}.
\label{lambda}
\end{eqnarray}
The different behaviours shown in Fig. \ref{tension} can be explained
in terms of the $n=1,2,3$ and $4$ components. 
Minima at $\simeq 45^{\circ}$ and
$135^{\circ}$ are due to the $n=4$ term. The $n=1$ and $3$ components affect
the relative stability of these two minima. The minimum at $90^{\circ}$
is due to the $n=2$ component. A model potential with no gap between
$\uparrow\uparrow$ and $\uparrow\downarrow$ configurations promotes a
$90^{\circ}$ angle at the boundary \cite{Yuri1} (planar orientation). Inplane 
dipoles also favour this configuration. The presence of a tilt-induced 
gap is responsible for the preferred orientations at oblique angles.

Competition between bulk interactions from perpendicular dipoles and an 
isotropic line tension $\lambda_0$ has been invoked as a mechanism
to determine domain shape in mesoscopic models. Dipole interactions
promote elongated shapes, which are counterbalanced by an isotropic 
line tension. Different shape regimes emerge from this competition.
However, the angle-dependent terms (\ref{lambda}), which are present 
even in the absence of dipolar interactions, play the same role as these interactions.
In this alternative scenario, long linear sectors of the domain boundary 
would optimise the line free energy, thus favouring the formation of
elongated domains with long linear boundaries. 
In fact, the results presented in this
section correspond to a value of the line-tension-to-dipolar strength
ratio of $\Lambda=\lambda_0^*/(\mu_{\perp}^*\Delta\rho^*)^2\simeq 0.02$, i.e. 
to a regime dominated by dipolar interactions. A reduction of $\mu_{\perp}^*$ to
$0.1$ would increase $\Lambda$ to $\sim 1$. Even in this case 
domains with a high number of lobes are stabilised, according to mesoscopic models
\cite{domains1,Aurora}. Anisotropic line tensions are expected to
suppress or at least reduce the stability of these highly-lobed structures in favour of
shapes with lower overall curvature.

\section*{Summary and conclusions}

In this work we have formulated a very simple model to study the interfacial structure
at the boundary between LC and LE domains in DPPC monolayers. The model predicts an
anisotropic line tension $\lambda(\varphi_0)$
with respect to the angle $\varphi_0$ between the nematic director (which
describes orientation of the projected molecular chains in LC domains) and the normal
to the boundary. The minimum line tension occurs at oblique
angles, which is in agreement with results from atomistic simulation \cite{us,Javanainen}.
Anisotropy in the line tension is already implicit in a model with only 
van der Waals lipid chain interactions (modified Gay-Berne potential), 
and dipolar components perpendicular to the monolayer only marginally 
increase
the anisotropy. Indirectly our model also supports the concept that inplane dipolar
components should have a minor role in the value of line tension since these components
favour a planar orientation at the boundary, which is incompatible with the results from
atomistic simulations. This may be explained by the stronger screening of
inplane dipoles as compared to perpendicular dipoles.

Our results indicate that theoretical mesoscopic models aimed at predicting domain shape and
domain shape transitions should be extended to account for anisotropy in the
line tension. On the one hand, inplane dipolar components may not be 
relevant to construct realistic models (see Ref. \cite{Moy} where the 
effect of such contributions is discussed).
On the other, models with  perpendicular dipoles should also 
reflect the anisotropy in the line tension stemming from 
the combined effect of nonzero tilt angles of lipid chains
and perpendicular dipoles. We should
recall that all atomistic simulations to date consistently
predict nonzero lipid-chain tilt angles in LC domains. In our model, the 
perpendicular dipole component seems to play an important role in reducing
the line tension, but only induces a small incremental anisotropy. 
Of course, competition between (long-range) dipolar bulk
interactions and line tension is a relevant factor for 
the global domain morphology,
as indicated by mesoscopic models \cite{domains1,domains2,Aurora}, but 
an anisotropic line tension may be a crucial requirement.  

\section*{Acknowledgments}

We acknowledge financial support from grants
FIS2017-86007-C3-1-P and FIS2017-86007-C3-2-P from Ministerio de
Econom\'{\i}a, Industria y Competitividad (MINECO) of Spain.

\newpage



\end{document}